\documentclass[conference]{IEEEtran}
\IEEEoverridecommandlockouts
\usepackage{amsmath,amsfonts}
\usepackage{tabularx}
\usepackage{subfig}
\usepackage{graphicx}
\usepackage{physics}
\usepackage{textcomp}
\usepackage{xcolor}
\usepackage{fancyhdr}
\usepackage{enumitem}
\usepackage{afterpage}
\usepackage{array}
\usepackage{bm}
\usepackage{float}
\usepackage[normalem]{ulem}
\usepackage{caption}
\usepackage{environ}
\usepackage{textcomp}
\usepackage{booktabs}
\usepackage{makecell}
\usepackage{tcolorbox,tikz}

\usepackage[noend]{algpseudocode}
\usepackage{algorithm}

\tcbuselibrary{skins}
\tcbuselibrary{breakable}
\newtcolorbox{mybox}[3][]
{
  breakable, 
  enhanced,
  colback         = #2!10,
  colframe        = #2!5,
  boxsep          =-0.5mm,
  borderline west = {1.0mm}{0.05mm}{#3!30}, 
  borderline north = {0.3mm}{0.05mm}{#3!30}, 
  borderline east = {0.3mm}{0.05mm}{#3!30}, 
  borderline south = {0.3mm}{0.05mm}{#3!30}, 
  #1,
}

\usepackage{pifont}% http://ctan.org/pkg/pifont
\newcolumntype{P}[1]{>{\centering\arraybackslash}p{#1}}
\colorlet{shadecolor}{gray!20}

\newcommand{\todo}[1]{\textcolor{black}{#1}}

\def\BibTeX{{\rm B\kern-.05em{\sc i\kern-.025em b}\kern-.08em
    T\kern-.1667em\lower.7ex\hbox{E}\kern-.125emX}}

\begin{document}
\bstctlcite{IEEEexample:BSTcontrol}

\author{
        % Primary authors
        Baolin Li\IEEEauthorrefmark{1},        
        Yankai Jiang\IEEEauthorrefmark{1},
        Vijay Gadepally\IEEEauthorrefmark{2},
        Devesh Tiwari\IEEEauthorrefmark{1}\\
    \IEEEauthorrefmark{1} Northeastern University,
    \IEEEauthorrefmark{2} MIT}

\title{LLM Inference Serving: Survey of Recent Advances and Opportunities}
% \subtitle{Paper Type: Open-source tools or data}


\maketitle

\begin{abstract}
This survey offers a comprehensive overview of recent advancements in Large Language Model (LLM) serving systems, focusing on research since the year 2023. We specifically examine system-level enhancements that improve performance and efficiency without altering the core LLM decoding mechanisms. By selecting and reviewing high-quality papers from prestigious ML and system venues, we highlight key innovations and practical considerations for deploying and scaling LLMs in real-world production environments. This survey serves as a valuable resource for LLM practitioners seeking to stay abreast of the latest developments in this rapidly evolving field.
\end{abstract}

\section{Introduction}
\label{sec:intro}

% Talk about LLM and deployment challenges

Large language models (LLMs) have rapidly gained immense popularity since the release of ChatGPT. However, deploying and scaling these powerful AI models in production environments has presented significant challenges. The substantial computational and memory demands of LLMs often necessitate the use of high-performance GPU servers, yet even these resources can be strained by the sheer size of the models and the lengthy text sequences they process.

The growing demand for LLM-powered applications has fueled a surge of research into LLM serving systems. In this paper, we present a comprehensive survey of these systems, focusing on advancements since 2023. While previous LLM system research existed, the landscape has dramatically shifted within the last year. Nearly every major system conference now features dedicated sessions on LLMs, with a particular emphasis on serving systems due to their widespread deployment and the importance of low-latency performance for user experience.

The sheer volume of research published in such a short timeframe makes it difficult for LLM practitioners to stay abreast of developments and identify the most promising approaches for real-world deployment. This survey aims to provide a clear overview of the current state of the art, highlighting key areas of innovation and practical considerations for production environments.

In this survey, we have meticulously selected all the high-quality research papers focused exclusively on LLM serving systems, published between January 2023 and June 2024. Our selection criteria prioritized publications from prestigious machine learning (ML) and system venues (e.g., ASPLOS, MLSys, OSDI), as well as impactful arXiv submissions from established industry and academic research groups. Notably, we exclude studies that modify LLM decoding algorithms (e.g., multiple decoding head~\cite{cai2024medusa}, lookahead decoding~\cite{fu2024break}, key token selection~\cite{adnan2024keyformer}) and solely focus on system-level enhancements that maintain the integrity of standard LLM decoding processes.

While a few prior LLM inference system surveys exist~\cite{miao2023towards,yuan2024llm,zhou2024survey}, these generally cover a broader scope and do not specifically emphasize system research. Additionally, many of the papers discussed in those surveys involve decoding algorithm modifications that can affect model accuracy. Our survey, in contrast, explicitly focuses on system-level solutions that do not alter the core LLM decoding mechanisms. Moreover, our survey encompasses a significant body of research published after the release of these earlier surveys, thus providing a more comprehensive and up-to-date overview of the field.

We have organized the recent advances in LLM serving systems into four distinct categories, each with its own set of challenges and opportunities, which we will delve into in the following sections.

\vspace{1mm}
\noindent\textbf{KV cache and memory management. } Efficient memory management is crucial to handle the dynamic growth of KV caches, which store previous key-value pairs to accelerate LLM inference. Recent research explores non-contiguous memory allocation, distributed management, and intelligent caching strategies to optimize memory utilization. Compression techniques are also being investigated to reduce the overall memory footprint, ultimately enhancing LLM performance and scalability by allowing for longer context lengths and lower memory overhead.

\vspace{1mm}
\noindent\textbf{LLM computation optimization. } Efforts to optimize LLM computation focus on request batching to maximize resource utilization. Additionally, disaggregating the inference process into prefill and decode phases enables independent optimization and hardware specialization. Model parallelism, employing various techniques, facilitates efficient execution across multiple GPUs. These strategies collectively enhance LLM execution efficiency and hardware utilization.

\vspace{1mm}
\noindent\textbf{Cloud LLM deployment. }Cloud platforms provide a scalable and cost-effective foundation for LLM inference. However, challenges remain in optimizing costs and resource utilization. Research is addressing this through techniques such as spot instance management, serverless optimizations, intelligent resource allocation, and power management. Additionally, strategies like cloud task co-location and token delivery optimization enhance user experience and overall cloud efficiency.

\vspace{1mm}
\noindent\textbf{Emerging research fields. } Emerging areas in LLM serving include retrieval-augmented generation (RAG) and mixture-of-experts (MoE) inference. RAG faces challenges related to the computational overhead of increased input lengths due to retrieved documents, while MoE inference grapples with efficient communication and load balancing between distributed experts. Other research efforts address ethical concerns in LLM serving, such as fairness and environmental sustainability, for which we provide a comprehensive list of relevant studies.

% carefully organized into various categories

% create an overview figure about the research directions

% create a timeline figure about the publications

% Discuss difference compared to previous survey
\section{Background}
\label{sec:bkgd}

\subsection{Overview of Transformer-based LLM Architecture}
Mainstream LLMs are built on multiple transformer blocks~\cite{vaswani2017attention}. Each identical transformer primarily consists of self-attention-based \emph{Multi-head Attention} (MHA) operations and \emph{Feed-Forward Networks} (FFN). Initially, the transformer applies three weight matrices ($W^Q, W^K, W^V$) to the input $X$ (encoded representation of input text sequence) to compute queries $Q$, keys $K$, and values $V$. Then, the \emph{Self-attention} is calculated as:

\vspace{-3mm}
{\small
\begin{align}
\begin{split}
 Q = XW^Q;K = XW^K;V = XW^V\\
    \text{Attention}(Q, K, V)=\text{softmax}(\frac{QK^T}{\sqrt{d_k}})V
     \notag
\end{split}
\end{align}}
\label{eq:softmax}
\vspace{-3mm}

This is the calculation of one attention head ($H_i$), and multiple heads are concatenated and linearly projected into the final attention result: 

\vspace{-3mm}
{\small
\begin{align}
\begin{split}
    H_i = \text{Attention}(XW^{Q}_{i},XW^{K}_{i},XW^{Q}_{i}) \\
    \text{Multi-Head Attention} = \text{Concat}(H_1, H_2,..., H_h) W^O
     \notag
\end{split}
\end{align}}
\label{eq:multi}
\vspace{-3mm}

MHA makes transformers focus on
different parts of the sequence in different representational spaces. Next, following the MHA block, the normalized output is fed into a position-wise FFN, which consists of two linear transformations with a ReLU activation.

\vspace{-5mm}
{\small
\begin{align}
\begin{split}
    \text{FFN}(x) = \text{max}(0, xW_1 + b_1)W_2 + b_2
     \notag
\end{split}
\end{align}}
\label{eq:ffn}
\vspace{-5mm}

\begin{figure}[t]
    \centering
        % \vspace{-5mm}
    \includegraphics[scale=0.36]{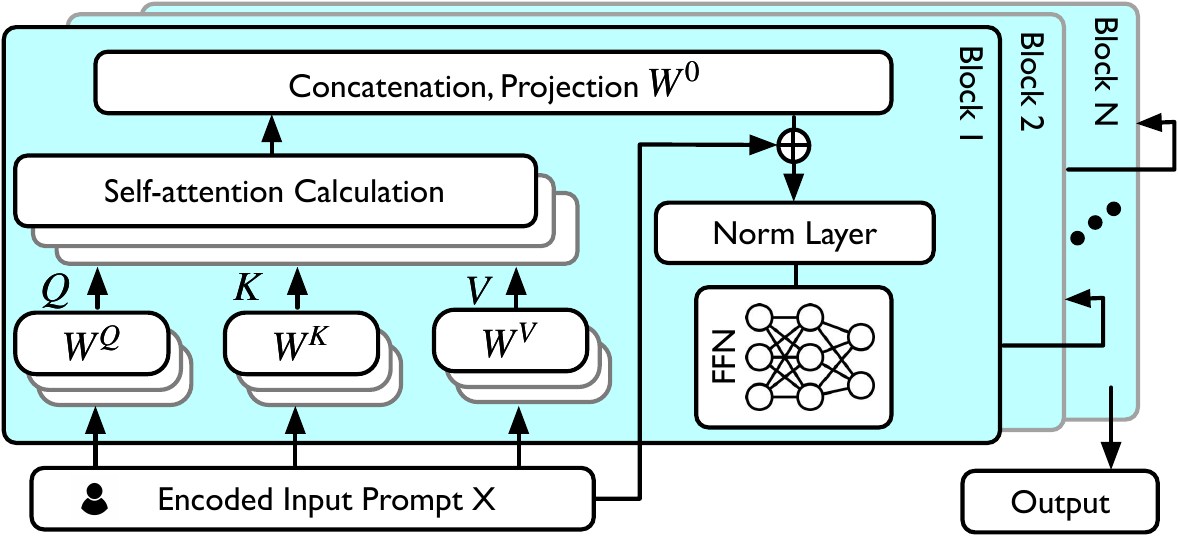}
    \vspace{1mm}
    \hrule
     \vspace{1mm}
    \caption{Transformer-based LLM architecture including both the multi-head attention mechanism and feed-forward network.}
    \vspace{-8mm}
    \label{fig:llm_arch}
\end{figure}

The FFN can be applied separately to each position, further refining the information captured by the MHA block. 
% Finally, \emph{Position Encoding} is added to give the model information about the sequence order of the tokens. 
The output will have the same dimension as the input $X$. Fig.~\ref{fig:llm_arch} provides a visualization of the LLM architecture.

\subsection{Overview of LLM Inference}
\begin{figure}[t]
    \centering
    \includegraphics[scale=0.4]{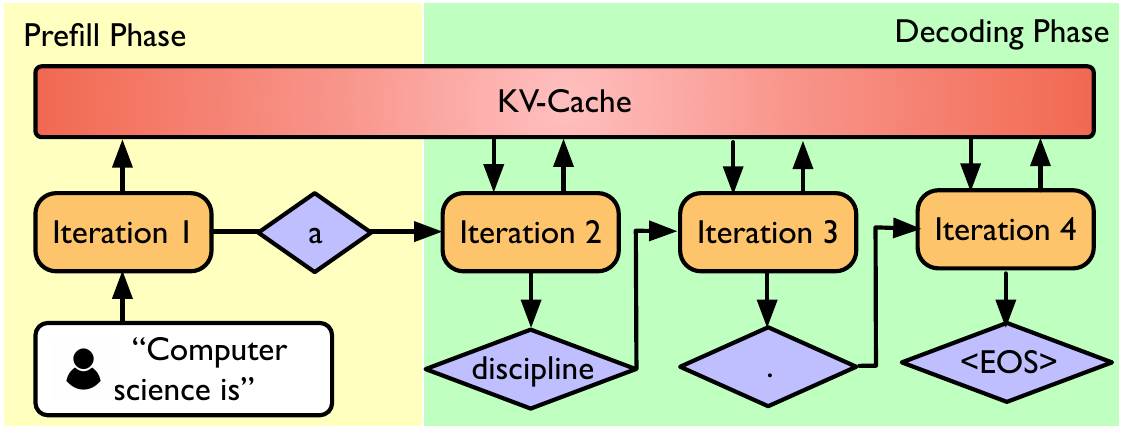}
    \vspace{1mm}
    \hrule
     \vspace{1mm}
    \caption{Prefill and decoding phase in the LLM inference.}
    \vspace{-8mm}
    \label{fig:llm_inference}
\end{figure}

LLM inference generates output tokens autoregressively~\cite{radford2018improving} based on the initial input sequences $P$, referred to as \emph{Prompts}. This process is divided into two major phases: the prefill phase and the decoding phase. The prefill phase is essential for setting up the model to generate text efficiently, while the decoding phase handles the generation of subsequent tokens. We visualize this process in Fig.~\ref{fig:llm_inference}.

The prefill phase starts with a tokenized and encoded representation of the prompt going through layers of the transformers. Note that the generated key-value ($KV$) pairs of all transformer blocks are cached during the prefill phase, referred to as KV cache~\cite{pope2023efficiently}. It ensures that the model can generate tokens more efficiently without recomputing the KV vectors of all previous tokens. Let the input prompt \( P = [p_{1}, p_{2}, ..., p_{n}] \), during the prefill phase, a new token is generated, denoted as \( P_{n+1} \), and the new $K$ and $V$ are cached as \( [(k_1,v_1), (k_2,v_2), ..., (k_n,v_n)] \).

The decoding phase is where the model generates new tokens autoregressively. The LLM predicts the next token, appends the newly generated token \( p_{n+1} \) to the original prompt \( P \), and updates the KV cache. Note that the KV cache grows linearly with the number of tokens generated. The autoregressive LLM inference process is outlined in Algorithm~\ref{alg:llm}.

\vspace{-2mm}
\begin{algorithm}
    \caption{Autoregressive LLM Inference}
    \begin{algorithmic}[1]
    \Statex \textbf{Input} $P$: encoded input sequence $[p_{1}, p_{2}, ..., p_{n}]$
    \Statex \textbf{Output} $X$: generated new sequence $[]$.
    \State Forward Pass ($[p_{1}, p_{2}, ..., p_{n}]$)
    \State Store the KV cache: $[(k_1,v_1), (k_2,v_2), ..., (k_n,v_n)]$
    \For {$i$ from 1 to M}
        \State Predict the next token $p_{n+i}$ using the KV cache.
        \State Store $(k_{n+i}, v_{n+i})$ to the KV cache.
        \State $X \gets X \cup \{p_{n+i}\}$
        \If{$p_{n+i}$ is EOS token \textbf{or} len($X$)$>$max length}
            \State \textbf{break}
        \EndIf
    \EndFor
    \end{algorithmic}
    \label{alg:llm}
\end{algorithm}

\section{Memory Management and Caching}
\label{sec:part1}

% Describe the challenges in the space, with a figure
% In this section, we investigate memory management techniques to reduce the memory footprint and memory access overhead during LLM inference. Overall, the inference memory footprint includes the LLM model parameters, the intermediate activations, and the KV cache. The model parameters remain constant during autoregressive generation, the activations only occupy a small fraction of GPU memory~\cite{kwon2023efficient}, while the KV cache grows larger and larger as more tokens are generated. As a result, LLM inference system research has focused on supporting efficient management of KV cache, in order to support larger batch size inference and longer context processing.

In this section, we explore memory management techniques to mitigate memory footprint and access overhead during LLM inference. While model parameters remain constant and intermediate activations are relatively small, the KV cache -- used to store attention information -- grows substantially with the number of generated tokens. Therefore, recent research has focused on efficient KV cache management to enable larger batch sizes and longer context processing.

\subsection{Efficient Management of KV Cache}

PagedAttention~\cite{kwon2023efficient} identifies that the KV cache dynamically grows and shrinks over time as the model generates new tokens, but the request generation lifetime and length are not known a priori. Thus, it proposes to manage the KV cache as non-contiguous memory blocks. Compared to contiguous KV cache, non-contiguous KV cache management significantly reduces the memory waste on pre-allocation and fragmentation. Due to its efficient memory management using pages, PagedAttention has become an industry norm in LLM serving frameworks, supported by TGI~\cite{tgi}, vLLM~\cite{kwon2023efficient} and TensorRT-LLM~\cite{tensorrtllm}. 

Despite its success, researchers still identify its weakness as PagedAttention requires rewriting attention kernels to accommodate the non-contiguous memory blocks, its memory manager adds software complexity and redundancy, and introduces performance overhead. Recently, vAttention~\cite{prabhu2024vattention} was proposed to retain the KV cache in contiguous virtual memory. It leverages pre-existing low-level system calls for demand paging, which is a standard operating system feature to reduce the software complexity. vAttention overlaps memory allocation with computation, pre-allocates memory ahead of time, and defers memory reclamation to hide the latency of memory allocation and improve the overall performance of the system.

Besides system memory management, other efforts have addressed application-specific KV cache efficiency. Prompt Cache~\cite{gim2024prompt} designs specific prompt schema for users to submit their requests, so that attention states from these pre-defined modules (e.g., system prompt) can be reused across multiple prompts. AttentionStore~\cite{gaocost} identifies that human interactions with applications such as ChatGPT are mostly multi-turn conversations. However, LLM engines would discard the KV cache when the user session becomes inactive to free up HBM space for other active sessions and re-compute the whole KV cache again when the session becomes active, leading to extra pre-filling costs. AttentionStore utilizes slower-mediums (e.g., CPU memory and disk), overlaps KV cache loading with computation, and designs intelligent pre-fetching and eviction policies.

\subsection{Support for Long-Context Applications}

Serving long-context LLM applications is particularly challenging as the size of the KV cache scales with the number of tokens. The limited memory limits LLM's ability to handle long sequences, demanding more memory-efficient solutions. Ring attention~\cite{liu2023ring} is a novel distributed approach that leverages blockwise computation of attention and feedforward of long sequences across multiple devices. It efficiently overlaps KV cache communication with computation and extends the context length by the device count times. Infinite-LLM~\cite{lin2024infinite} is another distributed solution, it breaks down KV cache into smaller manageable units called rBlocks across GPUs/CPUs, and efficiently manages them with dynamic memory sharing and coordination. MemServe~\cite{hu2024memserve} unifies handling of inter-request and intra-request optimizations for LLM serving by introducing MemPool, a distributed memory pool to manage KV cache across all cluster memory and employs a global scheduler to maximize KV cache reuse.

When the context grows larger than the GPU memory limit, most systems offload the KV cache to the CPU. InfiniGen~\cite{lee2024infinigen} is a solution that speculates the important KV cache entries by rehearsing the attention computation of the current layer in the preceding layer and prefetches only the essential entries to the GPU, thereby reducing the data transfer overhead. LoongServe~\cite{wu2024loongserve} introduces a new parallelism paradigm called Elastic Sequence Parallelism (ESP) to dynamically adapt to resource usage variance between requests and phases (pre-filling and decoding) of a request. It reduces KV cache migration overhead and KV cache fragmentation when serving long sequences.

\subsection{Compression of KV Cache}

Due to the large memory footprint of LLM serving, some systems have resorted to compressing the KV cache. On top of memory aggregation and communication scheduling, FlexGen~\cite{sheng2023flexgen} uses fine-grained groupwise quantization to compress the weights and KV cache to 4 bits. KIVI~\cite{liu2024kivi} analyzes the element distribution of the LLM KV cache and applies asymmetric quantization of the Key and Value cache. KIVI quantizes the key cache per-channel (grouping elements along the channel dimension) and the value cache per-token to achieve minimum quantization error. Gear~\cite{kang2024gear} achieves near-lossless high-ratio KV cache compression by quantizing the majority of entries of similar magnitudes and employs a low-rank matrix to approximate the quantization error. MiniCache~\cite{liu2024minicache} observes that the KV cache states exhibit high similarity between adjacent layers in the middle-to-deep portion of LLMs. Based on this insight, MiniCache leverages this high similarity to merge them into a shared representation to reduce redundancy, while also identifying and retaining distinct states that are crucial for maintaining the model's performance, preventing information loss during compression.
\section{Computation Task Scheduling}
\label{sec:part2}

% discuss challenges
Besides memory and KV cache management, the computation of LLM also presents significant system challenges. Due to the sequential dependency between tokens during the autoregressive generation, LLM can only generate one token at a time for each request. Thus, LLM inference workloads are less resource-efficient than training workloads on GPU hardware that is designed for massively parallel execution. Following this incentive, we investigate system solutions that optimize the scheduling of computation tasks during the inference process.

\subsection{Request Batching}

When a single request cannot efficiently utilize the GPU, it is intuitive to batch multiple inference requests together to boost the occupancy of GPU cores. However, as responses to different prompts can have significantly variable lengths, when batched together, the shorter responses are forced to wait for the longer ones to complete, resulting in computational waste. Response Length Perception and Sequence Scheduling~\cite{zheng2024response} instructs the LLM to predict the response length before starting to generate the actual response, and batches queries with similar predicted response lengths to reduce computational waste. A similar approach, $S^3$~\cite{jin2023s}, finetunes a Distillbert model for sequence length prediction. Upon mispredictions, it preempts sequences that exceed their allocated memory and retrain the predictor to learn from its mistakes.

Generation length prediction based batching is less practical due to the strong reliance on the predictor. Orca~\cite{yu2022orca} proposes continuous batching at the token level rather than the request level. It continuously schedules new requests into the batch as soon as a request in the current batch completes. Continuous batching now has become an industry standard in LLM serving frameworks, incorporated into the software of TGI, vLLM, and TensorRT-LLM. Based on continuous batching, DeepSpeed-FastGen~\cite{holmes2024deepspeed} proposes a dynamic SplitFuse mechanism that decomposes long prompts into smaller chunks scheduled across multiple iterations and composes short prompts together to maintain the inference running at high throughput region (bounded by GPU compute not memory bandwidth). A similar idea was explored in Sarathi-Serve~\cite{agrawal2024taming}, which splits prefill requests into smaller chunks and schedules them alongside ongoing decode requests without causing stalls (stall-free batching). This allows new requests to join a running batch without pausing ongoing decodes, leading to minimal pipeline bubbles.

\subsection{Disaggregated Inference}

LLM inference goes through a prefill stage to process the prompt, populate the KV cache, and start the decoding stage to generate tokens (Sec.~\ref{sec:bkgd}). Existing LLM serving systems colocate the two
phases and batch the computation of prefill and decoding
across all users and requests. However, these two phases display distinct characteristics and can interfere with each other when requests at the prefill stage are batched with requests at the decoding stage. TetriInfer~\cite{hu2024inference} separates prefill and decode instances, allowing each phase to run independently and preventing interference between batch-like prefill jobs and latency-critical decode tasks. It employs a two-level scheduling algorithm that incorporates predicted resource usage to avoid scheduling hotspots during the decode phase, ensuring efficient resource allocation and minimizing contention.

Splitwise~\cite{patel2023splitwise} extensively characterizes the differences in the execution and utilization patterns of the prefill and decoding stage on different generations of GPUs (heterogeneous hardware). Splitwise proposes to split these two phases into separate machines, allowing for specialized hardware for each phase to achieve better utilization, reduce hardware ownership costs, and save energy. DistServe~\cite{zhong2024distserve} designs a placement algorithm to schedule the prefill and decoding stage computation tasks. In clusters with high-speed cross-node networks, DistServe optimizes parallelism configurations for prefill and decoding instances independently to achieve the best per-GPU goodput; In clusters with limited cross-node bandwidth, it ensures that prefill and decoding instances of the same stage are co-located within a single node and optimizes parallelism configurations within the node.

\subsection{Model Parallelism}

LLMs can have hundreds of billions of parameters, requiring
model parallel execution on multiple GPUs. Pope et al.~\cite{pope2023efficiently} develop an analytical model for inference efficiency, enabling the selection of optimal multi-dimensional partitioning techniques tailored for TPU v4 slices based on specific application needs. HeteGen~\cite{xuanlei2024hetegen} introduces a framework for heterogeneous parallel computing using CPUs and GPUs. It employs a heterogeneous parallel computing algorithm to distribute computation within its hybrid heterogeneous parallelism framework and enables asynchronous overlap to mitigate I/O bottlenecks between the CPU and GPU.

ExeGPT~\cite{oh2024exegpt} can find an optimal schedule control variable of the batch size and tensor parallelism degree that maximizes inference throughput while adhering to a given latency limit. It leverages the distribution of input and output sequence lengths to allocate resources efficiently and determine the best parallelism configuration. Helix~\cite{mei2024helix} is designed to partition an LLM across heterogeneous GPUs and different types of network connections. It formulates its model partition scenario as a max-flow problem of a directed, weighted graph whose nodes represent GPU instances and edges capture both GPU and network heterogeneity through their capacities in the max-flow problem.

\section{LLMs in the Cloud}
\label{sec:part3}

LLM deployments are computationally intensive and often require significant infrastructure to run effectively. Cloud platforms offer a scalable and cost-effective solution for deploying LLMs, eliminating the need for expensive hardware investments. The flexibility of cloud deployment allows organizations to easily adjust resources as needed, ensuring optimal performance and minimizing downtime. However, the significant costs associated with cloud computing resources and the challenge of ensuring their efficient utilization can be major obstacles for LLM service providers.

\subsection{Cloud Deployment Cost}

Modern clouds offer a variety of spot instances (e.g., AWS EC2 Spot Instance, Azure Spot Virtual Machines, Google Cloud Spot VMs). These instances run on spare capacity and are offered at highly discounted prices, but may be preempted at any time when other instances need the capacity. SpotServe~\cite{miao2024spotserve} addresses the challenges of using these instances for LLM serving, such as how to quickly adapt to changes in available instances and how to minimize the cost of migrating instances when interruptions occur. It also introduces a stateful inference recovery mechanism for inference engines to commit their progress at the token level and efficiently resume interrupted requests.

% serverless
Serverless is a recently emerged cloud computing paradigm, where inference service users can submit their model to the cloud and the cloud provider takes care of all infrastructure provision and scaling with varying inference request load, and saves unused hardware costs for customers. A major challenge in serverless is mitigating cold start, where a service instance would be shut down after not being accessed for some time, and once a new request arrives, it would experience a latency spike associated with re-initializing the service instance. ServerlessLLM~\cite{fu2024serverlessllm} addresses these latency issues by utilizing the underutilized storage and memory resources available on GPU servers. It introduces a new checkpoint format and loading system to speed up LLM model loading, a live migration mechanism to avoid interrupting ongoing inferences, and a locality-aware server allocation strategy to minimize LLM inference cold start latency.

Cloud providers often offer a wide range of heterogeneous instance selections labeled at different prices. Mélange~\cite{griggs2024m} is a cloud resource allocation framework that considers three key LLM service characteristics: request size, request rate, and service-level objective. It automatically navigates through the GPU option space to determine the most cost-efficient heterogeneous GPU allocation for a given LLM service. With the resources allocated and model hosted on the GPUs, Llumnix~\cite{sun2024llumnix} is a dynamic scheduling system for LLM serving that addresses the challenges of heterogeneous and unpredictable requests by rescheduling them across multiple model instances at runtime -- similar to how OS context switches across cores. Llumnix introduces an efficient live migration mechanism for requests and their in-memory states, minimizing downtime during rescheduling, and employs a dynamic scheduling policy that unifies various rescheduling scenarios, such as load balancing, de-fragmentation, prioritization, and auto-scaling. This efficiency has resulted in significant cost savings while achieving similar tail latency.

\subsection{Cloud Efficiency}

A key bottleneck resource in cloud datacenters is power, which LLMs are quickly saturating due to their growing computation demand. POLCA~\cite{patel2024characterizing} characterizes the power consumption patterns of LLMs in the cloud and finds that while training LLMs demands a lot of power and can strain the data center's power infrastructure, inference tasks offer more flexibility for power management due to their less predictable power demands. POLCA devises a framework to manage power in LLM inference clusters by dynamically applying techniques such as GPU frequency locking and power capping. PerLLM~\cite{yang2024perllm} takes the LLM inference to an edge-cloud collaboration scenario, where it leverages the strengths of edge computing (low latency, reduced energy costs) and cloud computing (high processing power) to handle LLM inference tasks efficiently. PerLLM employs a Constraint Satisfaction Upper Confidence Bound (CS-UCB) algorithm to optimize service scheduling and resource allocation while adhering to constraints like processing time, bandwidth, and computing power -- achieving energy LLM efficiency. 

Workloads often get co-located in the cloud environment. FlexLLM~\cite{miao2024flexllm} is a system designed to efficiently service LLM inference and parameter-efficient fine-tuning (PEFT) requests in the same iteration. LLM inference, which involves generating text token by token, is primarily limited by memory bandwidth due to the need to access all model parameters for each token generation. In contrast, PEFT, which processes all tokens of a request simultaneously, is mainly constrained by compute resources, such as the tensor cores on GPUs. FlexLLM introduces a token-level fine-tuning mechanism that breaks down the fine-tuning process into smaller, more manageable token-level computations to minimize memory usage and inference latency, making co-serving feasible. 

As LLM inference follows token-by-token generation, users also read the response word-by-word. Andes~\cite{liu2024andes} defines a user experience metric of Quality of Experience (QoE) for text streaming services. It is formulated by comparing the actual token delivery timeline (TDT) of a request with its expected TDT. The expected TDT is determined by the expected time to first token (TTFT) and the expected token delivery speed (TDS), which can vary depending on factors like the user's typical reading speed. The intuition is generating text too fast (than user reading speed) does not yield QoE benefits, wasting cloud resources. Andes addresses this by strategically allocating GPU resources among multiple requests to optimize QoE. It employs a dynamic priority-based preemptive scheduler that operates at the token level, prioritizing urgent requests and preempting those that have been sufficiently served. Andes improves average QoE and can handle higher request rates while maintaining similar token generation throughput.

\section{Emerging Research Fields}
\label{sec:part4}

\subsection{Retrieval Augmented Generation}

Retrieval-Augmented Generation (RAG)~\cite{lewis2020retrieval} is a technique that enhances LLMs by incorporating external information sources. It addresses the limitations of LLMs in retaining factual knowledge and their tendency to generate inaccurate or fabricated information (hallucinations). RAG operates in two stages: retrieval and generation. During retrieval, the system identifies the most relevant contexts from an external knowledge base or corpus based on the given query. Once the relevant contexts are retrieved, they are integrated into the LLM's generation process in different processes including concatenation (where the retrieved contexts are simply appended to the query) and cross-attention (where the LLM attends to the retrieved contexts during generation).

Sparse RAG~\cite{zhu2024accelerating} observes that RAG can be computationally expensive due to the increased input length from retrieved documents. It first encodes retrieved documents in parallel to eliminate latency caused by long-range attention, then selectively decodes the output by attending only to highly relevant caches chosen via prompting the LLM with special control tokens. RAGCache~\cite{jin2024ragcache} caches intermediate states of external knowledge with a knowledge tree to organize and store intermediate states. The cached knowledge can be shared across multiple queries to reduce the redundant computation. Another knowledge caching technique is CacheBlend~\cite{yao2024cacheblend}, which selectively recomputes a small portion of the KV cache based on the preceding text in the input. 

\subsection{Mixture-of-Experts Inference}

The mixture of Experts (MoE) is used in LLMs to improve efficiency and performance. It divides the model into specialized sub-networks, called ``experts", each focusing on a specific task. A ``gating" network then directs input to the most suitable expert. In the inference process of an MoE transformer, the input is first passed through a gating network. This network determines which expert, or a combination of experts, is best suited to process the specific input. MoE's sparsely activated subset of experts avoids the large computational need to process the entire model for every inference.

\vspace{1mm}
\noindent\textbf{MoE Communication. } Lina~\cite{li2023accelerating} is a system designed to address the all-to-all communication bottleneck in distributed MoE. The all-to-all communication occurs when distributed MoE sends tokens to their selected experts for processing and then sends the results back to the original devices. During inference, Lina dynamically schedules resources based on expert popularity, balancing the transfer size and bandwidth of all-to-all communication across devices. ExFlow~\cite{yao2024exploiting} is an optimization technique to accelerate the inference of distributed MoE. It leverages the inter-layer expert affinity, which is the correlation between expert selection across different MoE layers. By placing experts on corresponding GPUs based on their affinity, ExFlow reduces cross-GPU routing latency and improves inference throughput. 

\vspace{1mm}
\noindent\textbf{Expert offloading. } SiDA-MoE~\cite{du2024sida} (Sparsity-inspired Data-Aware) leverages both main memory and GPU memory by exploiting the inherent sparsity of expert activation in MoE models. SiDA-MoE includes two parallel threads: an inference thread and a hash-building thread. The hash-building thread predicts which experts will be activated for each token at each layer, storing these predictions in a hash table. The inference thread then uses this information to dynamically load activated experts onto the GPU and offload inactive experts to main memory, maximizing GPU memory utilization. MoE-Infinity~\cite{xue2024moe} takes a different approach toward expert offloading. The system leverages the observation that MoE models exhibit sparse activation and temporal locality during inference, meaning only a few experts are repeatedly activated for processing a specific sequence. MoE-Infinity traces expert activation at the sequence level, enabling it to predict which experts will be needed and prefetch them accordingly. 

\vspace{1mm}
\noindent\textbf{MoE Efficiency. } Fiddler~\cite{kamahori2024fiddler} is a system designed to efficiently run these models on a limited number of GPUs, even when the model's size would typically exceed the GPU's memory capacity. Fiddler strategically distributes the model's components. Non-expert layers, which are used frequently, are kept on the GPU. A subset of expert layers, chosen based on how often they're used, are also placed on the GPU. The rest remain in the CPU's memory. Huang et al.~\cite{huang2023towards} introduce three optimization techniques to address the MoE inference inefficiencies. (i) Dynamic gating allows the number of tokens processed by each expert to vary, which avoids the over-provisioning of resources in static gating and reduces computational waste, communication overhead, and memory consumption. (ii) Expert buffering leverages the observation that expert activation is often sparse and exhibits temporal locality. By caching frequently used (hot) experts in GPU memory and buffering less active experts in CPU memory, expert buffering reduces the static memory allocation on GPU. (iii) Imbalanced token assignments to experts can lead to bottlenecks and performance degradation. Expert load balancing ensures a more even distribution of workload across devices.

\subsection{Miscellaneous Fields} 

\noindent\textbf{Ethics and environmental sustainability. } Sheng et. al~\cite{sheng2023fairness} ensure fairness in serving LLMs by introducing a Virtual Token Counter (VTC). VTC defines LLM serving fairness based on a cost function that accounts for the number of input and output tokens processed. It achieves fairness by tracking the services received by each client and prioritizing those with the least service, while also considering the varying costs of processing input and output tokens. Sprout~\cite{li2024toward} addresses the environmental sustainability of LLMs and designs a framework to reduce the carbon footprint of LLM inference services. Sprout introduces ``generation directives" to guide the autoregressive generation process, balancing the need for sustainability with the demand for high-quality generation.

\vspace{1mm}
\noindent\textbf{Inference pipeline optimization. } FlashDecoding++~\cite{hong2024flashdecoding++} conducts inference engine performance optimization, addressing several issues in softmax synchronization, GPU kernel, and dataflow. For example, the decoding phase performs linear GEMM operations with flat shapes where the batch size dimension involved in the multiplication is much smaller than the others. FlashDecoding++ accelerates flat GEMM with double buffering that overlaps computation and data transfer and hides the memory latency in loading input matrices. Parrot~\cite{lin2024parrot} is designed to optimize the performance of LLM-based applications that involve multiple LLM requests with complex workflows. Parrot performs data flow analysis and uncovers correlations across multiple LLM requests, and introduces a series of optimizations to improve performance. FlashAttention-3~\cite{shah2024flashattention} is a method to speed up attention for large language models and long-context applications. It introduces techniques like warp specialization and asynchronous block-wise operations to optimize GPU utilization. FlashAttention-3 achieves significant speedup on Hopper GPUs compared to its predecessor and reduces numerical errors in FP8 computations.

\vspace{1mm}
\noindent\textbf{Frugal inference. } FrugalGPT~\cite{chen2023frugalgpt} proposes several solutions to reduce the inference cost, such as prompt caching and LLM cascading which uses a sequence of LLMs, starting with cheaper ones and moving to more expensive ones only if necessary. SpecInfer~\cite{miao2024specinfer} applies speculative decoding using smaller, speculative models to predict the LLM's output, reducing the computational resources. These predictions are organized into a tree structure, and their accuracy is verified in parallel against the LLM. RouteLLM~\cite{ong2024routellm} dynamically selects between a stronger and a weaker LLM during inference to optimize the balance between cost and response quality. 

% \subsection{Carbon Footptint and Sustainability}
% \input{sections/discussions.tex}
\section{Conclusion}
\label{sec:conclude}

This survey has presented a comprehensive overview of recent advancements in LLM serving systems, emphasizing the importance of system-level solutions for enhancing performance and efficiency. We have highlighted key innovations for deploying and scaling LLMs, paving the way for the future development of LLM serving systems.

\todo{
\section*{Acknowledgments}
This material is based upon work supported by the Assistant Secretary of Defense for Research and Engineering under Air Force Contract No. FA8702-15-D-0001, and United States Air Force Research Laboratory Cooperative Agreement Number FA8750-19-2-1000. Any opinions, findings, conclusions, or recommendations expressed in this material are those of the author(s) and do not necessarily reflect the views of the Assistant Secretary of Defense for Research and Engineering, or the United States Air Force. The U.S. Government is authorized to reproduce and distribute reprints for Government purposes notwithstanding any copyright notation herein. 
}

\bibliographystyle{IEEEtran}
\bibliography{refs}

\end{document}